# THE RISK DISTRIBUTION CURVE AND ITS DERIVATIVES

Ralph Stern
Cardiovascular Medicine
University of Michigan
Ann Arbor, Michigan

stern@umich.edu

ABSTRACT

Risk stratification is most directly and informatively summarized as a risk distribution curve. From this curve the ROC curve, predictiveness curve, and other curves depicting risk stratification can be derived, demonstrating that they present similar information. A mathematical expression for the ROC curve AUC is derived which clarifies how this measure of discrimination quantifies the overlap between patients who have and don't have events. This expression is used to define the positive correlation between the dispersion of the risk distribution curve and the ROC curve AUC. As more disperse risk distributions and greater separation between patients with and without events characterize superior risk stratification, the ROC curve AUC provides useful information.

Many statistical methods of risk stratification have been developed for clinical risk assessment. Evaluation of the clinical utility of these methods requires risk estimation for the members of a patient population. The most direct and informative way to present these results and understand the achieved risk stratification is a risk distribution curve.[1] Previously Huang et al. [2] had proposed use of the predictiveness curve which also presents risk distribution. Neither approach is being used. Instead the ROC (receiver operating characteristics) curve is usually presented and the area under this curve evaluated. However the value of this approach has been questioned.[3,4] In this paper, the relationship between the multiple graphical presentations of risk stratification used in the literature is outlined, providing a better understanding of the value of ROC curve analysis.

**Lognormal example of a risk distribution curve**

The risk distribution curve of cardiovascular risk in adults has not been presented. The expected distribution when risk factors interact multiplicatively is lognormal.[5] For cardiovascular risk, this is especially so for two reasons. First, levels of individual risk factors, such as blood pressure and cholesterol, are often lognormal. This reflects the fact that their levels in turn are determined by the action of multiple other influences.[6] Second, an exponential increase in risk with increasing levels of a risk factor will also produce lognormal risk distribution curves.

A lognormal curve is one whose natural logarithm is normally distributed with mean µ and variance $\sigma^2$. Lognormal variates are greater than zero and lognormal curves are variably skewed to right.  The equation for a lognormal risk distribution curve where risk, r, varies from >0 to 1 is:

$$f(r) = \frac{1}{r\sigma\sqrt{2\pi}} e^{-\frac{(r-\mu)^2}{2\sigma^2}}$$

The mean and variance of a lognormal distribution are:

$$\text{mean} = e^{\mu + \frac{\sigma^2}{2}}$$

$$\text{variance} = e^{2\mu + \sigma^2}(e^{\sigma^2} - 1)$$

We have used published data from an analysis of the NHANES study to estimate the parameters of a lognormal curve describing the distribution of 10-year cardiovascular risk determined by the Framingham risk equations in US adults without coronary artery disease or risk equivalents.[7]  These are $\mu = -2.9248$ and $\sigma = 0.68830$.  Figure 1 shows this lognormal risk distribution curve in the upper left corner.  The frequencies are arbitrary values which provide an area under the curve of 1.

**Derivatives of the risk distribution curve.**

Diamond [8] showed that the frequency of patients with and without events is simply obtained by multiplying the frequency of individuals at any given level of risk by the risk and (1-risk), respectively. If 100 individuals are at 25% risk, 25 patients with events and 75 patients without events are expected. The graph to the right of the risk distribution curve presents these two derived risk distribution curves. The population risk is 6.8% so the area under the curve of patients with events is 0.068 and the area under the curve of patients without events is 0.932.

The relationship between these two curves is more readily appreciated by adjusting the area under each of these curves to 1, which is shown in the next graph to the right.

Finally, the ROC curve is derived from these two curves and shown in the upper right corner of Figure 1. At each level of risk, the area under the two curves above that level of risk is determined and used as the coordinates for the points on the ROC curve. The x coordinate is the fraction of patients without events above that level of risk (1-specificity or false positive rate) and the y coordinate is the fraction of patients with events above that level of risk (sensitivity or true positive rate).

A related curve is derived by presenting the fraction of the population along the x-axis, instead of the fraction of patients without events. This presentation has been used in the genetics literature [9] and is shown in the lower right corner of Figure 1.

Alternatively, the cumulative risk distribution curve can be derived from the risk distribution curve. This is shown in the lower left corner of Figure 1.

If the axes of the cumulative risk distribution curve are exchanged, the predictiveness curve [2] is obtained and this shown in the graph to the right.

Since each of the curves can be derived from the risk distribution curve, it is clear they all present information contained in the risk distribution curve.

**Relationship between dispersion of the risk distribution and discrimination**

Figure 2 shows two other lognormal risk distribution curves with the same mean risk of 0.068, but different dispersion. The narrow curve (σ of 0.1) assigns all patients a risk close to the mean. As a consequence the separation between higher and lower risk patients or between patients who have and don't have events is minimal, i.e. there is poor discrimination. Risk stratification methods generating narrow curves provide poor risk stratification and have less clinical utility. The broad curve (σ of 1) assigns a wide range of risks to patients. The much greater separation and discrimination, exceeding that displayed in Figure 1, provides superior risk stratification and would be more useful clinically.

**Relationship between the risk distribution and ROC curve AUC**

As the ROC curve represents a parametric curve, the area under the curve [10] is:

$$\int_0^1 FractionPatientsWithoutEvents(r)[\int_r^1 FractionPatientsWithEvents(x)dx]dr$$

where x is a dummy variable to avoid using r both in the integration limit and the integrand. This is equivalent to the equation derived by Pepe [11] by a different approach.

After substitution, the expression for the ROC curve AUC becomes:

$$\int_0^1 \frac{(1-r)f(r)}{1-r_{mean}}[\int_r^1 \frac{xf(x)}{r_{mean}}dx]dr$$

where $r_{mean}$ is the mean or population risk.

Inspection of the first equation makes it clear how the ROC curve AUC measures discrimination. The first term is the area under the risk distribution curve for patients without events. The term in brackets is a quantitative measure of the overlap between the risk distribution curves of patients with and without events. It is the fraction of the area under the risk distribution curve for patients with events above a given level of risk. Were there no overlap of the two curves, it would be equal to 1. In that case the outer integral is then simply the area under a frequency distribution curve, i.e. 1. If the two risk distribution curves were superimposable, the term in brackets would decrease from 1 to 0 across the risk distribution curves. The integral would then be 1/2 the area under a frequency distribution curve, i.e. 0.5.

**Relationship between dispersion of the risk distribution and ROC curve AUC**

The more disperse the risk distribution curve, the greater the area under the ROC curve. This is shown in Figure 3a, where the ROC curve AUC is shown for lognormal risk distribution curves with the same mean as in Figure 1, but values of σ up to 1. This depicts the risk stratification of the adult population by methods differing in discrimination, e.g. because they differ in the risk factors included. The ROC curve AUC increases linearly. Values of σ above 1 produce a plateau and then decline in AUC as the continuous lognormal distribution begins assigning patients risks above 1. Figure 3b shows the AUC's for the same range of σ, but expressed as a function of variance.

The ROC curve AUC for the narrow distribution of figure 2 is 0.530, for the broad distribution of Figure 1 is 0.700, and for the very broad distribution of Figure 2 is 0.765.

**Relationship between dispersion of the risk distribution and risk categorization**

Janes [4] and Pencina [12] prefer separating patients into categories using clinically defined risk thresholds. Broader distributions will classify more patients as high and low risk giving results that must reflect dispersion. The fraction of patients at low risk (0 to 0.05) is 0.001, 0.459, and 0.576 in the narrow ($\sigma=0.1$), broad ($\sigma=0.68830$), and very broad ($\sigma=1$) risk distributions, respectively, discussed above, while the fraction of patients at high risk (>0.2) is 0.000, 0.028, and 0.056 in the same distributions.

**Discussion**

Once accuracy or calibration for a risk stratification method is established, discrimination must be assessed. Any of the graphical methods of Figure 1 could be used, but the important point that a more disperse risk distribution represents superior discrimination is best appreciated from the risk distribution curve.

The key point that broader risk distributions characterize superior risk stratification methods has been made before [1,2,9,13] Harrell [13] wrote: "The worth of a model can be judged by how far it goes out on a limb while still maintaining calibration."

The area under the ROC curve is a commonly employed measure of discrimination. It was initially utilized to evaluate the performance of diagnostic tests, but has since been widely applied to prognostic evaluations. The ROC curve itself does not allow the underlying risk distribution to be visualized, which has obscured the rationale for using the ROC curve AUC. As more disperse risk distributions and greater separation between patients with and without events characterizes superior risk stratification, the ROC curve AUC provides useful information.

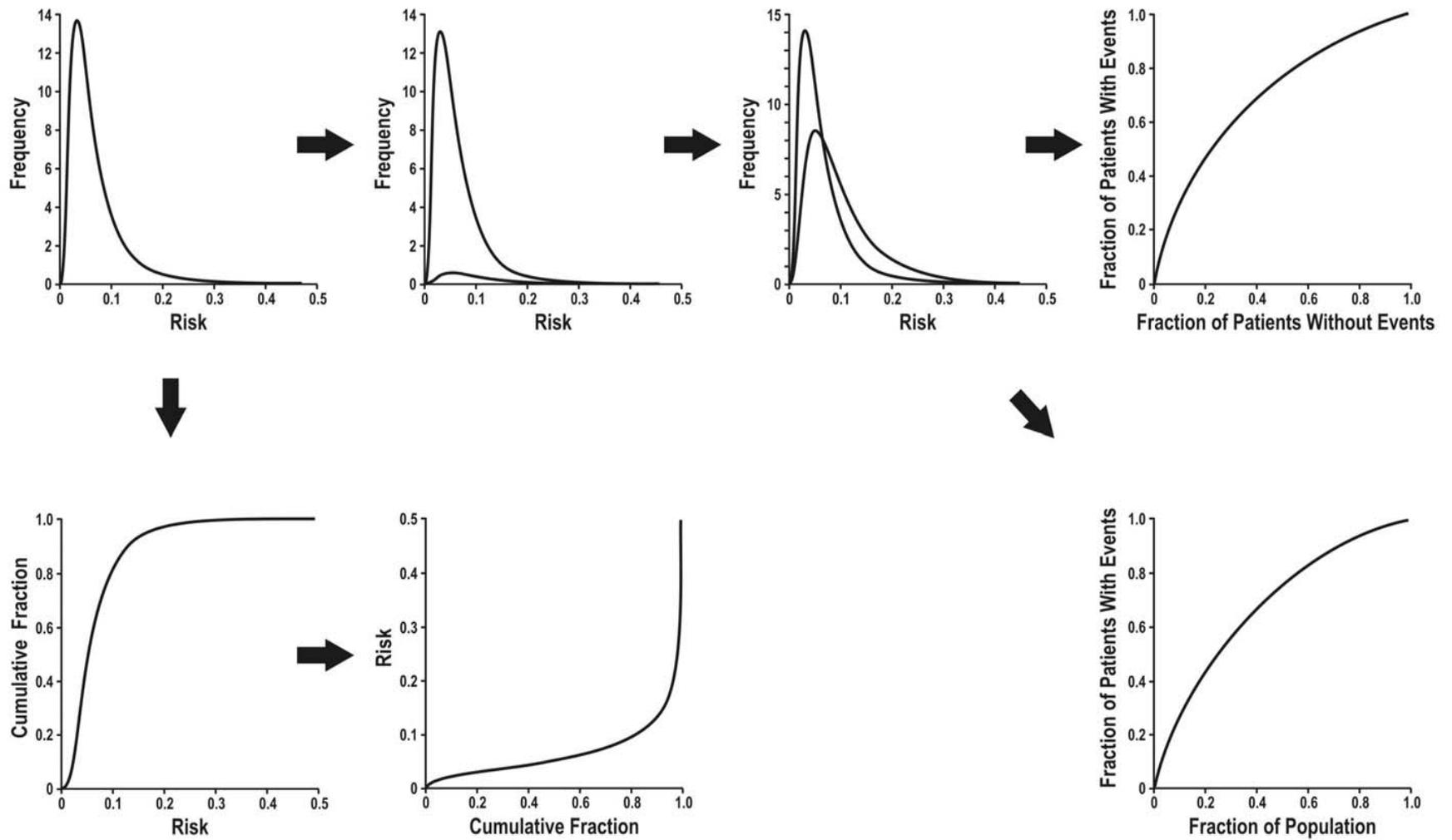

Figure 1. A lognormal risk distribution curve and its derivatives

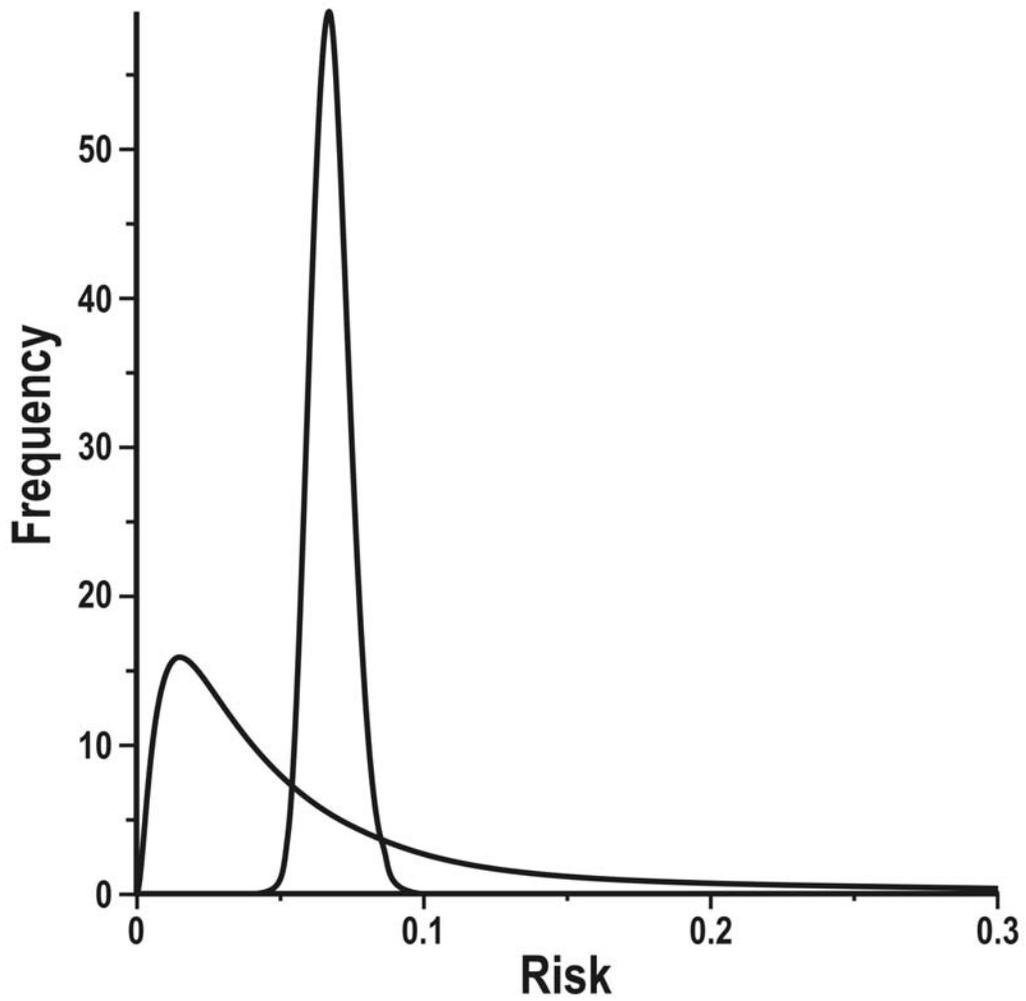

Figure 2. Lognormal risk distribution curves with the same means (0.068) but different dispersions (σ's of 0.1 and 1)

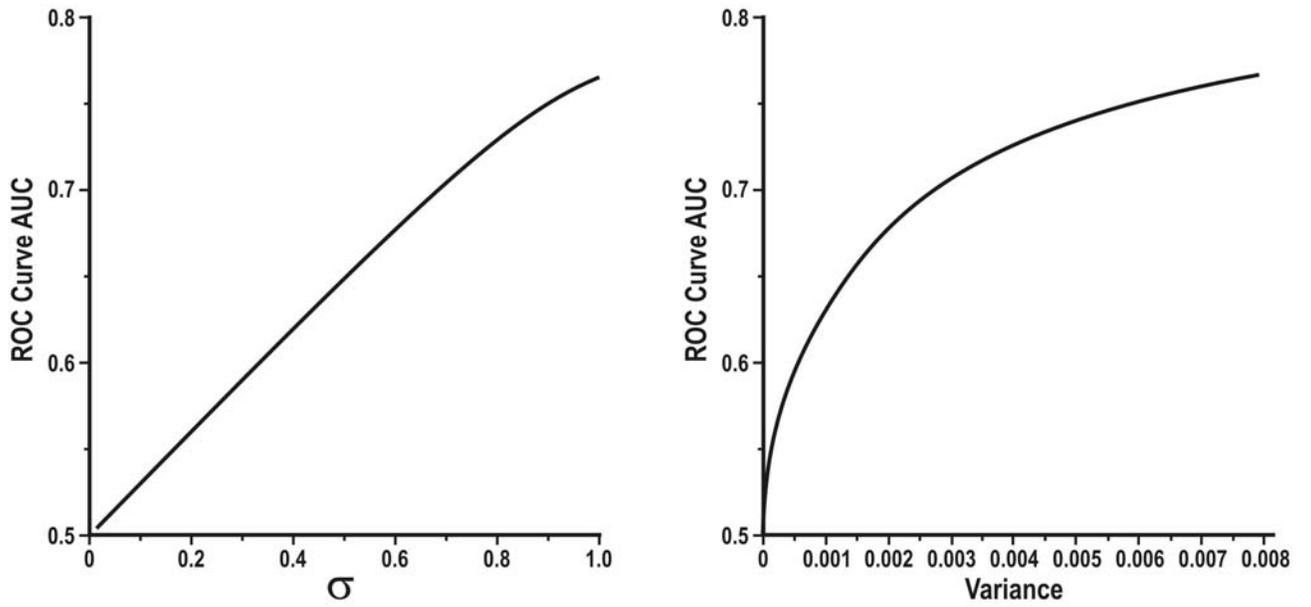

Figure 3. The relationship between ROC curve AUC and σ (left) and variance (right) for lognormal risk distribution curves with a mean of 0.068